%% file: hd165052.tex
\documentclass[usegraphicx,usedcolumn,usenatbib,epsf,graphics]{mn2e}
\usepackage{rotating}
\newif\ifAMStwofonts
\AMStwofontstrue

\ifCUPmtlplainloaded \else
  \ifAMStwofonts \else 

  \fi
\fi

\title[The massive double-lined O-type binary HD 165052]{The massive double-lined O-type binary HD~165052}

\author[Arias et al.]
{
J.I. Arias$^1$\thanks{Fellow of the Universidad Nacional de La Plata, Argentina},
N.I. Morrell$^1$\thanks{Member of Carrera del Investigador Cient\'{\i}fico, CONICET, Argentina},
R.H. Barb\'a$^1${\Huge$^{\dagger}$},
G.L. Bosch$^1$\thanks {Post-doctoral fellow of CONICET, Argentina},
M. Grosso$^2$\thanks {Member of the Carrera del T\'ecnico of CONICET, Argentina},
M. Corcoran$^{3}$\\
$^1$ Facultad de Ciencias Astron\'omicas y Geof\'{\i}sicas, Universidad Nacional de La Plata, Paseo del Bosque S/N, B1900FWA La Plata, Argentina\\
$^2$ Complejo Astron\'omico El Leoncito, Av. Espa\~na 1512 sur, 5400 San Juan, 
Argentina\\
$^3$ Universities Space Research Association, 7501 Forbes
Blvd, Ste 206, Seabrook, MD 20706, USA \& \\
Laboratory for High Energy Astrophysics, Goddard Space Flight Center, 
  Greenbelt MD 20771, USA\\
}

\pubyear{2001}

\begin{document} 

\maketitle

\begin{abstract}
We present a new optical spectroscopic study of the O-type binary HD~165052
based on high- and intermediate-resolution CCD observations. 
We re-investigated the spectral classification
of the binary components, obtaining spectral types of O6.5\,V and O7.5\,V
for the primary and secondary, respectively, finding that both stars
display weak C\,{\sc iii} $\lambda5696$ emission in their spectra.
We also determined a radial-velocity orbit for HD~165052
with a period of 2.95510 $\pm$ 0.00001 d, and semiamplitudes of
94.8 and 104.7 $\pm$ 0.5 km s$^{-1}$, resulting in a mass ratio $Q = 0.9$.
From a comparison with previous radial-velocity determinations, we
found evidence of apsidal motion in the system. 
Several signatures of wind-wind collision, such as phase-locked variability 
of the X-ray flux and the Struve-Sahade effect, are also considered.
It was also found that the reddening in the region should be normal,
in contrast with previous determinations.
\end{abstract}

\begin{keywords}
stars: binaries -- stars: early-type -- stars: individual: HD~165052 --
 X-rays: stars
\end{keywords}

\section{Introduction}
NGC~6530 is a very young open cluster located in the central part of 
the H\,{\sc ii} region M\,8 (the Lagoon Nebula, which contains also 
the NGC 6523 
nebula). NGC\,6530 is one of the more studied clusters in our
Galaxy. It has an interesting history of star formation which 
apparently is still in progress (Sung, Chun \& Bessel 2000). 

One of the O-type members of NGC~6530 is the double-lined binary 
HD\,165052 (= HIP~88581 = CD-24~13864; 
$\alpha_{2000}=18^{\rm h}05^{\rm m}11^{\rm s}$, 
$\delta_{2000}=-24^\circ 23' 54''$; $V=6.87$).
This star is refered as $\#118$ in the pioneering work of Walker (1957)
on this young open cluster.
The first indication of variable radial velocity in HD\,165052
was given by Plaskett (1924), while its double-lined binary nature was
pointed out by Conti (1974).
The first orbital solution for this binary system was
obtained by  Morrison \& Conti (1978), based on 
20 photographic spectrograms at a reciprocal dispersion of 17 \AA \,mm$^{-1}$. 
HD\,165052 was classified as O7: by Conti (1974) and O6.5\,V + O6.5\,V
by Morrison \& Conti (1978).
Because of the relatively modest resolution of their observations in addition
 to the remarkable
similarities between the two binary  components, 
in terms of both spectral type and 
linewidths, they were unable to  unambiguously distinguish the two
components in any individual observation. 
This led them to derive a period of 6.14 d for the system,
essentially doubling the true value.
Later, Stickland, Lloyd \& Koch (1997), showed that the real period was 
close to three days, based on 
15 high-resolution {\em IUE} spectra, some of them obtained during
four consecutive days.

Regarding the spectral classification of HD~165052, Walborn (1972) 
derived a spectral type O6.5\,V, later modified to  O6.5\,V(n)((f))
by the same author (Walborn 1973). 
The designation ((f)) indicates the presence of  very weak 
N\,{\sc iii} $\lambda$4634-40-42 \AA\ emission. 
As a result of their significant resemblance, the same spectral type 
has been often adopted for both binary components,
i.e. the system has been classified as O6.5\,V + O6.5\,V, 
and more recently, O6\,V + O6\,V (Penny 1996).

Further interest in this system arises from the fact that it is observed as 
an X-ray source.
X-ray emission in early-type binary systems appears as a natural consequence 
of stellar wind-wind interactions (Chlebowski \& Garmany 1991).
In such a case, the observed emission is 
expected to present orbital-phase-related variations.
The analysis of such phenomena obviously requires good knowledge 
of the orbital parameters, 
and it is important since it provides valuable information about the mass-loss 
rates and the terminal velocities of the stellar winds.
Corcoran (1996) analyzed the X-ray emission from hot, massive stars.
HD~165052 was one of the selected targets in which he detected 
the presence of
phase-dependent variations probably associated with  wind-wind collision.

\input{runs.tex}

Another expected consequence of colliding winds is the detection of 
emission in the H$\alpha$ line. 
A search for H$\alpha$ emission  in early-type binary systems 
recently performed  by Thaller (1997) yielded negative results for HD~165052. 

In this paper we present a new spectroscopic investigation of HD~165052 
based on high- and intermediate-resolution CCD observations.
From high-resolution optical echelle CCD spectra
we have determined  the radial-velocity
orbit of HD\,165052. We have also used those echelle CCD observations and
additional intermediate-resolution, high signal-to-noise Cassegrain CCD
 spectra to review the spectral classification of the
binary components of HD\,165052. Spectral-type variations
related to the Struve-Sahade effect are also described.

\section{Observations}

All the observations used in the present study were obtained at the 
Complejo Astron\'omico El Leoncito (CASLEO
\footnote{CASLEO is operated under agreement between CONICET and the 
National Universities of La Plata, C\'ordoba and San Juan}),
Argentina, with the 2.15-m Jorge Sahade Telescope between June 1994 and 
June 2001. 

The whole data-set contains 73 CCD spectra of HD~165052 
with different dispersions and spectral ranges (see Table~\ref{runs.tex}).

The  Boller \& Chivens (B\&C) spectrograph was used with a Thomson 
$400\times592$ CCD as detector, and a 1200 l~mm$^{-1}$ diffraction 
grating in its second order, yielding a reciprocal dispersion
of 0.6\,\AA\,px$^{-1}$ and a resolution $R=\lambda / \Delta(\lambda)$ of 3500. 
The approximate wavelength ranges covered with this 
configuration were 3900 to 4200\,\AA \,and 4400 to 4700\,\AA, in two different
grating angles. A He-Ar lamp was used as comparison source.

For the \'echelle spectra obtained 
with the modified REOSC SEL 
\footnote{Spectrograph Echelle Li\`ege (jointly built by REOSC and
Li\`ege Observatory and on long term loan from the latter).}
Cassegrain spectrograph in crossed-dispersion mode (REOSC-CD), a Tek 
1024 pixel CCD was used as detector.
The reciprocal dispersion of these data is 0.17\,\AA\,px$^{-1}$ at 
4500\,\AA\,and they have $R \sim 15000$.  
45 of them cover the approximate wavelength region 3800 to 6400\,\AA, 
whereas  five cover the region from 5200 to 7600\,\AA.
The comparison spectra were obtained with a Th-Ar lamp.

For the spectra secured with the REOSC SEL spectrograph in simple
dispersion mode (REOSC-SD),  the same detector was used
with a 600 l\,mm$^{-1}$ diffraction grating, this combination 
yielding a reciprocal dispersion 
of 2.5\,\AA\,px$^{-1}$ and a resolution $R\sim1800$, 
and covering wavelength regions 
from 3900 to 5500\,\AA \,and from 5400 to 7000\,\AA \,for two different
grating positions. A Cu-Ar lamp was used as comparison source for this
instrumental configuration.

The usual sets of bias, flat-fields and darks were also secured for each
observing night. 
The data were reduced and analyzed at La Plata Observatory with
 IRAF\footnote{IRAF 
is distributed by NOAO, operated by AURA, Inc., under agreement with NSF.} 
standard routines.

\section{Orbital Elements and their Discussion}

\subsection{Radial Velocities}

In order to determine an accurate radial-velocity orbit for the HD~165052 
binary system, we used only the high-resolution \'echelle CCD observations.
The high signal-to-noise ratio ($S/N$) of our \'echelle data 
allowed the detection of slight
spectral differences between the binary components 
that will be discussed in section~\ref{st}.

The radial velocities were measured by interactive fitting of
Gaussian profiles to the observed absorption lines.
Depending on the degree of blending, we applied simple or simultaneous
Gaussian fitting to the absorption-line profiles.
For the radial-velocity orbit calculation we considered only those
spectra in which the lines of both binary components were resolved.

He\,{\sc i} absorption lines, in particular $\lambda$$\lambda$\,3819, 4026, 
4471, 4921, 5015 and 5875, showed well defined profiles 
and they were measured in almost all cases. 
Being intrinsically  wider, He\,{\sc ii} $\lambda$$\lambda$\,4200, 4542, 
4686 and 5411 absorption lines were taken into account only
when the separation between binary components was  maximum. 
Other lines measured for radial velocities were Si\,{\sc iv} $\lambda$\,4088, 
O{\sc iii} $\lambda$\,5592 and C\,{\sc iv}  $\lambda$$\lambda$\,5801, 5812.
Table \ref{table2} shows the complete set of lines measured in
each spectrum. 
For each Julian date, the first and the second lines list the 
heliocentric radial velocities assigned from the different spectral 
lines to the primary and secondary components, respectively. 
There is only one set of measurements for single-lined spectra, 
which are presumably blended and consequently were omitted from the orbital
solution.
All the radial velocities are expressed in km\,s$^{-1}$.
The values labeled as $V_R$ (column 19) were computed as the simple 
average of the heliocentric radial velocities for the selected lines.
The corresponding standard deviations for these averages 
are shown in column 20.
Table \ref{vr} shows the final observed radial velocities for 
the spectra considered in the orbital solution. 
The column labeled as $n$ shows the number of lines averaged in each case.
The orbital phases listed in Table \ref{vr} were computed 
with the  ephemeris presented in Table~\ref{orb_sol}.

\input{radvel_all.tex} 

\input{vr3.tex}        

\subsection {The radial-velocity orbit}

\begin{figure*}
\includegraphics[width=150mm, bb= 95 310 480 510]{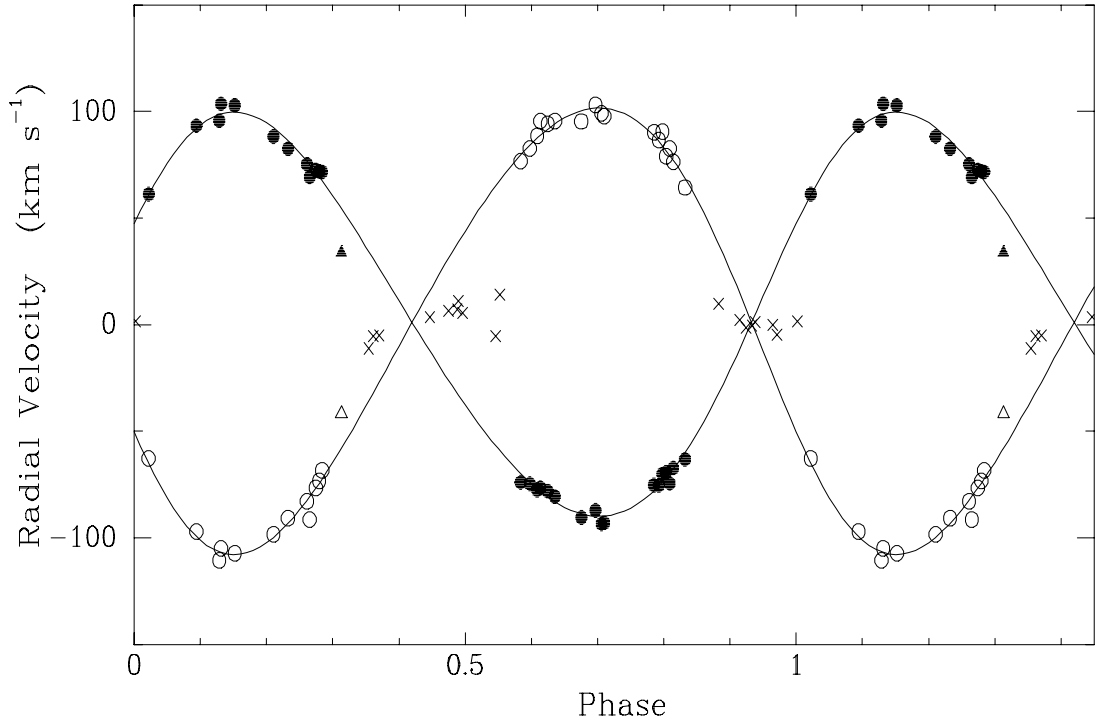}
\vspace{2cm}
\caption{The spectroscopic orbit of HD~165052 determined from high resolution
CASLEO observations. Filled and open symbols represent the primary and 
 secondary components, respectively. 
Triangles denote measurements not considered in the solution because of
pair blending and crosses depicts data where no double lines are observed.}
\label{orbita}
\end{figure*}

\input{orb_sol.tex}

To determine an improved orbital period for the HD\,165052 binary
system we used all the available radial velocities of this star, i.e., those
derived from our new high-resolution CCD observations, those
obtained by Morrison \& Conti (1978) from photographic spectra, which
were reassigned in accordance with the new period,
 and those derived by Stickland et al. (1997) from
{\em IUE} observations.  
The applied period-finding routines, which comprised the 
Lafler \& Kinman (1965) 
method and several subsequent modifications of it (e.g. Marraco \& Muzzio 
1980 and Cincotta, M\'endez \& N\'u\~nez 1995), 
led to a most probable value of $P=2.95510 \pm 0.00002$ days,
very close to the value determined by Stickland et al. (1997),
namely $2.95506 \pm 0.00001$.

The orbital elements were obtained from the new high-resolution radial
velocities only,
using a modified version 
of the code originally written by Bertiau \& Grobben (1969),
considering  simultaneously the radial-velocity measurements for 
both binary components.
One of the 30 double-lined spectra was discarded from the final
solution because of the pair blending effect.

\begin{figure*}
\begin{minipage}{180mm}
\hskip 10mm
\includegraphics[width=80mm, bb= 110 255 490 515]{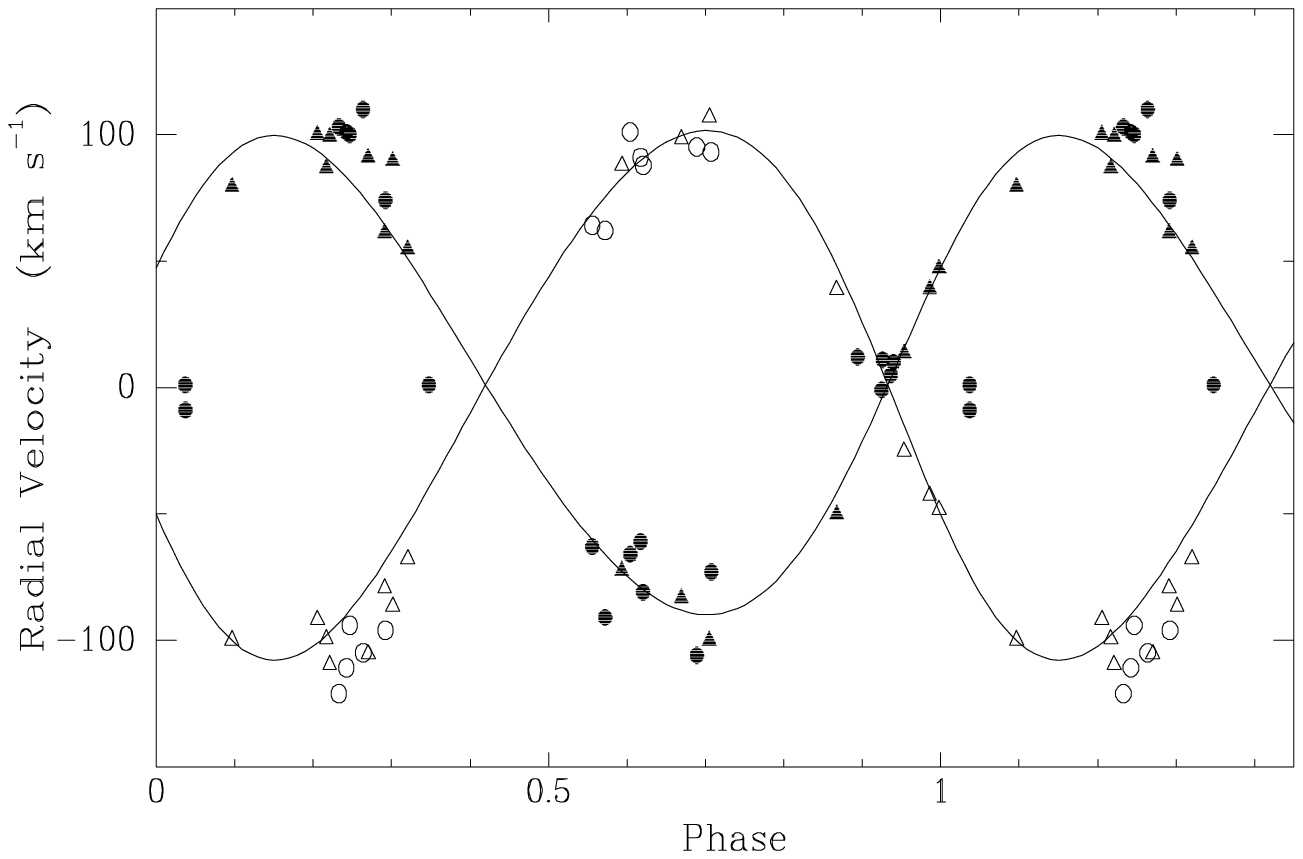}
\includegraphics[width=80mm, bb= 110 255 490 515]{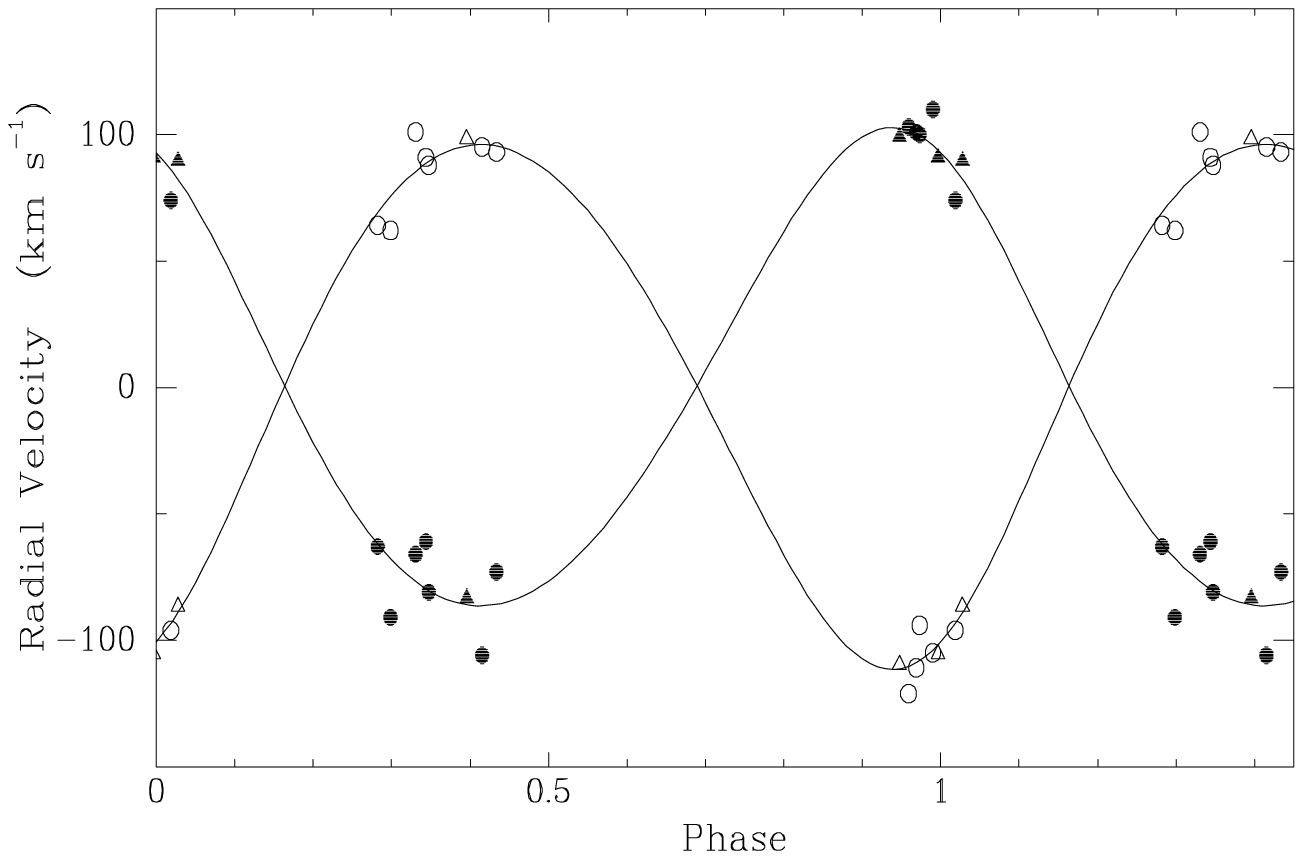}
\caption{Left-hand panel: previously published data in comparison with the
 spectroscopic 
orbit computed from our observations.
Filled and open triangles represent radial velocities derived from
$IUE$ data (Stickland et al. 1997) for the primary and secondary 
components, respectively. 
Circles represent the optical velocities from
Morrison \& Conti (1978). Right-hand panel: orbital solution computed from
the observations previous to 1982, adopting the period and 
eccentricity obtained in this work. Filled symbols represent the 
primary component and open symbols the secondary one, 
with circles denoting optical and squares IUE velocities.}
\label{apsides}
\end{minipage}
\end{figure*}

The resulting orbital elements are listed in Table~\ref{orb_sol}.
We have found a slight eccentricity, $e=0.09$, which is almost
certainly real since trial solutions with the eccentricity fixed to 0.0 yield 
systematic shifts between the fitted curve and the observations, and the 
probable error derived is $\sim 5$\,km\,s$^{-1}$, more than double 
the error of the adopted eccentric orbit fit.
Moreover, in accordance with the criterion developed by Lucy \& Sweeney (1971)
for deciding between an elliptical and a circular orbit, 
the eccentricity will be signficant at the $5\,\%$ level 
if it exceeds 3.63 times its probable error, which is widely satisfied  
in this case.

Figure~\ref{orbita} shows the observed radial velocities
along with the orbital solutions of Table~\ref{orb_sol}. 

Figure \ref{apsides} (left-hand panel) 
shows the radial-velocity curves determined 
from our high-resolution data in comparison with previous observations 
(Morrison \& Conti 1978 and Stickland et al. 1997).
The significant shift between the orbital solution
 obtained from the new data set and the previous observations 
might be an indication of apsidal motion.
Apsidal motion changes the shape of the radial-velocity of a binary 
curve when observed at different epochs, as described, for 
example, by Morrell et al. (2001) for the orbit of HD~93205.  
The precise knowledge of apsidal rotation in a binary system is important 
as it may provide a way to find an independent estimate of the masses of 
the binary components (see Benvenuto et al. 2002).
Consequently, we decided to investigate this hypothesis, calculating an 
orbital solution for a data subset containing the oldest observations, 
i.e., previous to 1982, adopting the values of 
$P = 2.95510$~days and $e=0.09$ computed in this work for
the orbital period and the eccentricity. 

As a result, we found that the old observations fit  
the new orbit derived from high-resolution echelle spectra, 
if one only varies the longitude of periastron $\omega$.  
The value $\omega$ that best describes the binary motion between 1973 
and 1982 is $\omega=26\fdg4\pm19\fdg6$ 
instead of $\omega=296\fdg7\pm3\fdg5$ resulting from the 
new calculation. This solution is illustrated in  
Figure \ref{apsides} (right-hand panel).
The probable error of this fit is only 6.3\,km\,s$^{-1}$.

\begin{figure}
\includegraphics[width=90mm]{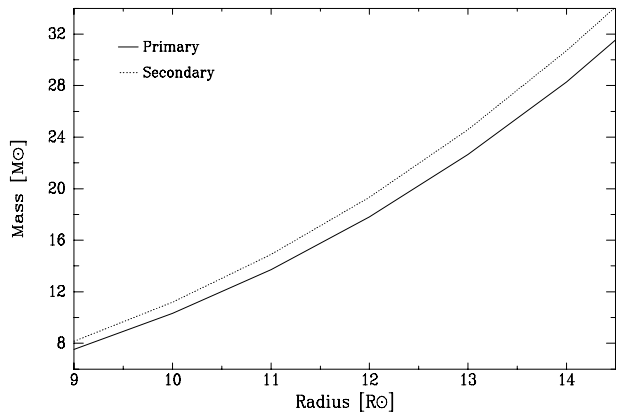}
\caption{Mass-radius relationship for HD~165052 assuming 
synchronously rotating stars.}
\label{masas}
\end{figure}

With the above-determined orbital solution and taking the estimates for
$V_e \sin i$ found from {\it IUE} spectra by Stickland et al. (1997),
i.e., 85 and 80 km\,s$^{-1}$ for the primary and secondary components 
respectively,
we derived a mass-radius relationship for a system of two synchronously
rotating stars.
The resulting curves,
$M_1=0.010309 \times R_1^3$ and $M_2=0.011189 \times R_2^3$,
are shown in Figure~\ref{masas}.
As the binary components are both dwarfs (see next section for details), 
one expects individual radii of $\sim10~R_{\odot}$ 
(Vacca et al. 1996),
which would indicate masses close to $10 M_{\odot}$ for each component.
These values are obviously too low for O-type stars, and  
the reason for this inconsistency is probably 
the hypothesis of synchronization.
Mass estimates for early O-type stars are still controversial.
If the  masses are in the expected range for the corresponding 
spectral types (30-40 $M_{\odot}$), the orbital inclination would be
around $i=20^\circ$, leading to rotational velocities near 250 km\,s$^{-1}$. 
This would imply rotation periods of about 2 d in both cases.
It is worthwhile to mention that the {\it Hipparcos} 
photometry shows no evidence 
of variability in excess of 0.01 magnitude with the orbital period,
in good agreement with the expectations for the system's small inclination 
and under-Roche filling stellar radii.

\section{Spectral classification and emission lines}
\label{st}

\begin{figure}
\includegraphics[width=90mm]{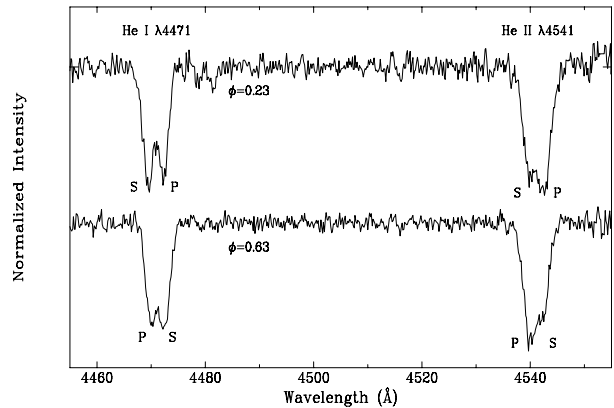}
\caption{Rectified spectrograms of HD~165052 in the region of 
the He\,{\sc i} 4471\,\AA\ and He\,{\sc ii} 4542\,\AA\  
absorption lines at nearly opposite orbital phases. 
``P'' and ``S''  refer to the primary and secondary components,
respectively.}
\label{spec}
\end{figure}

HD~165052 has often been considered 
as a system composed  of two stars identical in spectral type, i.e.
 O6.5\,V + O6.5\,V or more recently O6\,V + O6\,V (Penny 1996), 
but what emerged from the analysis of our high-resolution CCD spectra 
is that one of the components (the primary in mass terms) is earlier than 
the other, although the difference corresponds to no more than one 
spectral subclass.
Thus, the primary component shows a spectrum in which the ratio 
He\,{\sc ii} $\lambda4542$/He\,{\sc i} $\lambda 4471$ is larger than unity 
and corresponds to the spectral type O6.5\,V.
On the other hand, in the secondary spectrum the He\,{\sc i} absorption lines
are slightly stronger than the He\,{\sc ii} lines, 
and the ratio He\,{\sc ii} $\lambda4542$/He\,{\sc i} $\lambda 4471$ 
is somewhat lower than unity, indicating  a spectral type O7.5\,V.
Figure~\ref{spec} displays the region containing the 
He\,{\sc i} 4471\,\AA\ and He\,{\sc ii} 4542\,\AA\ absorption 
lines in two high-resolution \'echelle spectrograms at  
opposite phases of the binary period. Also apparent in Figure~\ref{spec} is
a phase-dependent change in the intensity of the absorption lines of
the secondary component which will be discussed in the next section.

\begin{figure*}
\includegraphics[width=85mm]{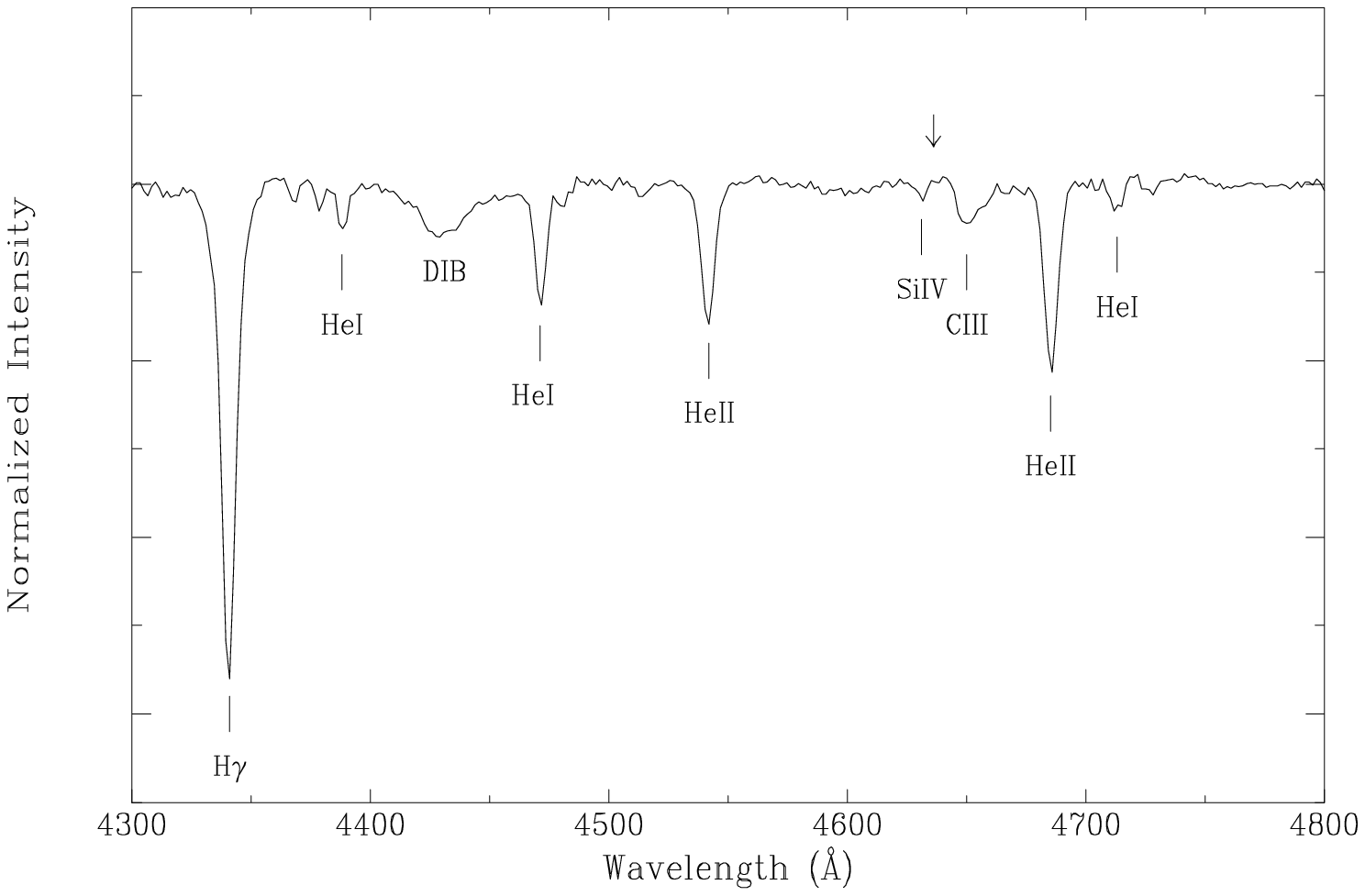}
\includegraphics[width=85mm]{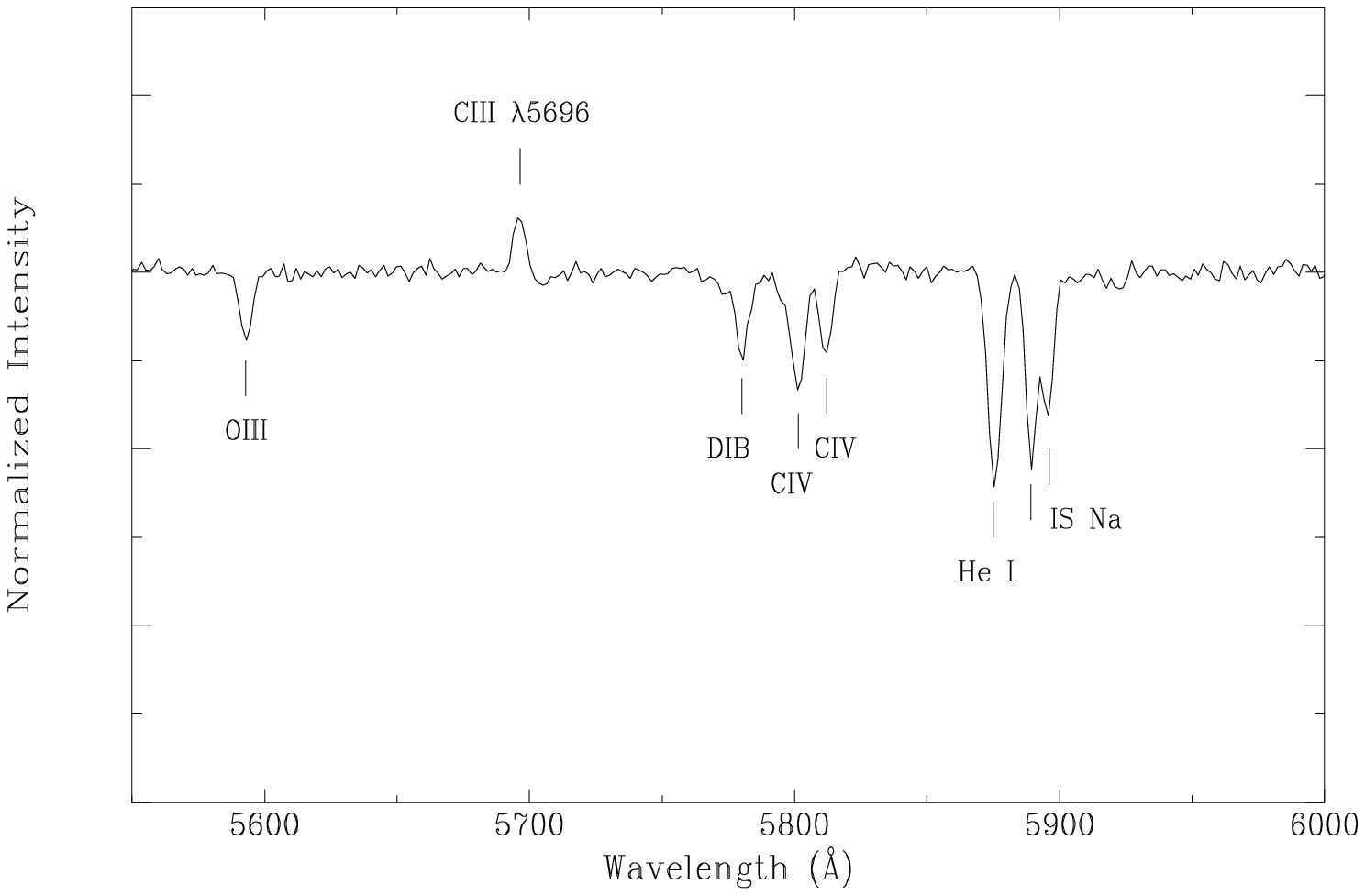}
\caption{Low-dispersion spectra of HD~165052.  
N\,{\sc iii} $\lambda$$\lambda$4634-40-42 emissions should be 
observed where the arrow indicates (left-hand panel). 
C\,{\sc iii} $\lambda$5696 emission is clearly observed above the continuum
(right-hand panel).}
\label{emissions}
\end{figure*}

According to the O-type  optical luminosity criteria 
(Walborn \& Fitzpatrick 1990), 
at spectral classes O6 to O8, one finds 
He\,{\sc ii} $\lambda$4686\,\AA\ in strong absorption on the main sequence; 
in the intermediate luminosity classes  
this absortion line weakens and may become neutralized;
and finally, the O supergiants have this feature strongly in emission.
A correlative increase with luminosity in the strengths of the 
Si\,{\sc iv} absorption lines flanking H$\delta$ can also be seen at these
spectral types.
We find both phenomena in the spectrum of HD~165052
(see Figure~\ref{emissions} later),
indicating that it must be a class V system.

We also searched the {\em IUE} database and retrieved all the available
observations of HD~165052, along with those corresponding to HD~190864
and HD~93146 presented as O6.5\,III(f) and O6.5\,V((f)) prototypes 
respectively, in the 
atlas of ultraviolet spectra by Walborn, Nichols-Bohlin \& Panek (1985). 

\begin{figure*}
\includegraphics[width=80mm]{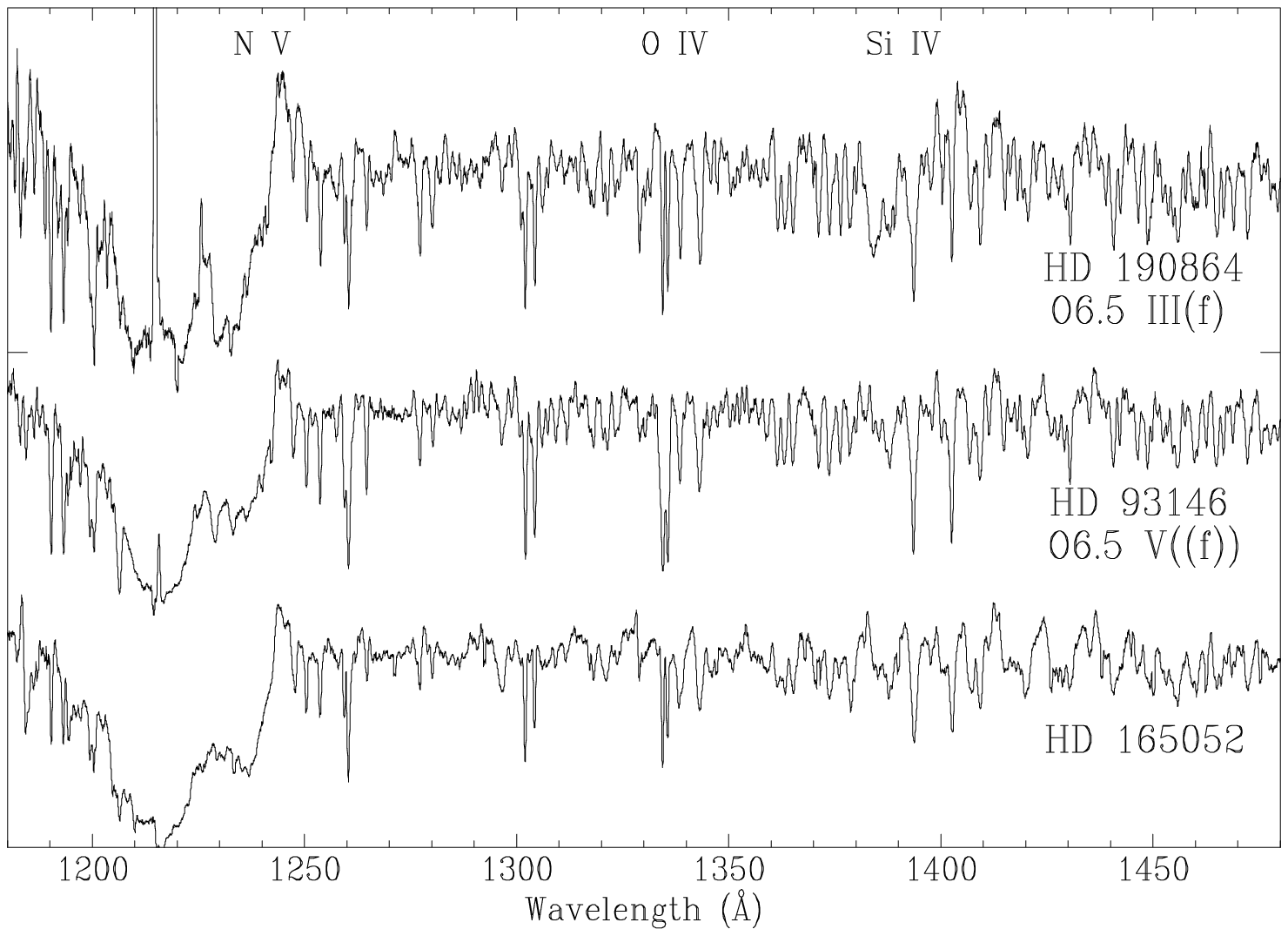}
\includegraphics[width=80mm]{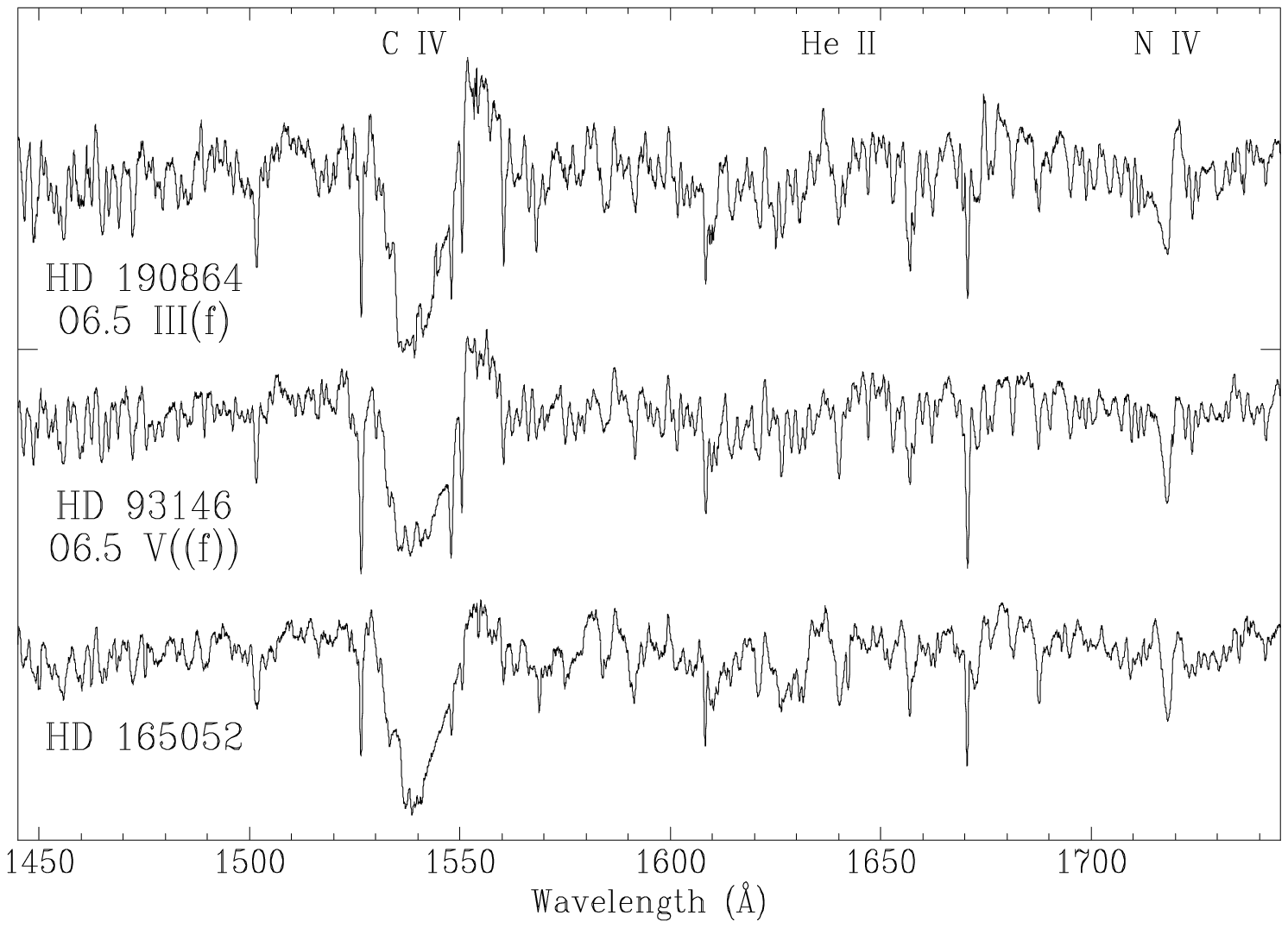}
\caption{{\em IUE} spectra of HD~165052 and the classification
standards HD~190864 and HD~93146. A thorough comparison among them
brings more evidence in support of the luminosity class V for the binary
components.}
\label{iue}
\end{figure*}

A simple inspection of those ultraviolet spectra showed that 
the stellar-wind profile of the Si\,{\sc iv} doublet at $1394-1403$ \,\AA, 
which displays a pronounced luminosity dependence with no effect at class V
and adquiring a P Cygni type profile at higher luminosities,
is completely absent in the spectrum of HD~165052. The same
occurs with the profile of N\,{\sc iv}\,1720 \,\AA, which shows a full
P Cyg profile at III(f) but no trace of winds in the dwarf spectra. 
Moreover, the stellar-wind profiles of the doublets 
N\,{\sc v} $1239-1243$ \,\AA\ and C\,{\sc iv} $1548-1551$\AA\ 
are much less developed in the spectrum
of HD~165052 than in the spectrum of the giant HD~190864
(see Figure~\ref{iue}).  
These considerations add more evidence in support of 
the classification of HD~165052 as a class\,V system. 

An interesting point concerns the reddening in the region.
McCall, Richter \& Visvanathan (1990) found the 
anomalous value of $R=4.6 \pm 0.3$ for the reddening law in the
Lagoon Nebula, and it is apparently uniform throughout the region.
However, this value of $R$ leads to an absolute visual 
magnitude $M_V=-5.71$ for each binary component (see Stickland et al. 1997), 
which would be appropiate for luminosity class III objects
(Vacca et al. 1996; Schmidt-Kaler 1982).
On the other hand, adopting a normal value of $R=3.2$, 
together with the 
color excess $E(B-V)=0.43$ estimated for HD~165052 and a distance modulus of
$11\fm25$ (Sung et al. 2000), one finds for the binary system $M_V=-5.76$. 
Assuming that both components contribute with equal flux, this implies 
$M_V=-5.0$ for each one, 
in fair accord with the values assigned to O6.5\,V stars in the above
mentioned calibrations.

\begin{figure}
\includegraphics[width=80mm]{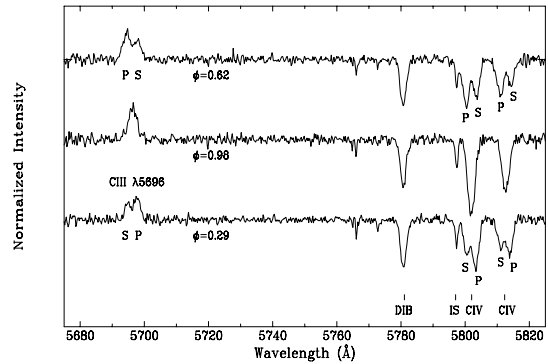}
\caption{Rectified spectrograms of HD~165052 at three different orbital
phases showing the duplicity of the C\,{\sc iii} $\lambda$5696 emission line. 
The C\,{\sc iv} absorption lines $\lambda$5801 and $\lambda$5812 are 
also displayed in this figure.}
\label{double}
\end{figure}

Many of the massive stars also show certain emission lines related 
to luminosity. 
This characteristic is usually identified by adding the designation 
``f'' to the spectral type.
Main-sequence spectra in the range O6-O7 present strong 
He\,{\sc ii} $\lambda$4686\,\AA\ absorption and often 
weak N\,{\sc iii} $\lambda$$\lambda$4634-40-42 emission, 
which corresponds to the notation 
((f)) according to Walborn's (1971) classification criteria. 
In addition, one frequently finds a very weak C\,{\sc iii} 
$\lambda$5696 emission in types from O4 to O8 (Walborn 1980).
These emissions being quite weak, their detection is strongly affected 
by observational parameters. 
The new CCD observations of HD~165052, with $S/N \sim 300$ but 
$\lambda/\Delta(\lambda)$ only 1800, 
do not show the N\,{\sc iii}~$\lambda$$\lambda$~4634-40-42 emission 
reported by Walborn (1973) (see Figure~\ref{emissions}).
However, after careful inspection of the corresponding photographic
observation obtained at Cerro Tololo Inter-American Observatory in 1972,
which was kindly provided by Dr. Walborn, we are able to confirm the existence 
of those emission lines,
although very weak, in the spectrum of HD~165052.
On the other hand, 
the presence of C\,{\sc iii} $\lambda$5696 emission 
is shown in Figure~\ref{emissions}. 
From the analysis of the complete set of 
\'echelle spectra, we discovered that this feature is actually a 
double line which shares the binary motion of both components of HD~165052, 
thus demonstrating that the C\,{\sc iii} emission 
arises in both stars and not only in the primary as was suggested 
by Conti (1974)  (Figure~\ref{double}). Also, the primary component 
of this emission line is stronger than the secondary.
Also evident from Figure~\ref{emissions} is the strength of the He\,{\sc ii} 
4686\,\AA\ absorption, indicating an early evolutionary state
(main sequence or zero age main sequence)  for this system.

\section{Wind-wind Interaction}

Massive luminous stars are known to be sources of strong stellar winds. 
In early close binary systems the individual stellar winds may collide 
in a bow shock zone between the stars.
This idea is strongly supported by the growing observational evidence
from investigations of H$\alpha$ (Thaller 1997), UV wind lines 
(Koch et al. 1996)
and X-ray emission properties (Corcoran 1996; Pittard \& Stevens 1997)
in binaries containing O-type stars.

Since HD~165052 is a short-period binary with luminous components, it is a 
good candidate to show wind-wind collision effects. 

One observational consequence of the presence of wind interactions in a 
massive binary is the detection of H$\alpha$ emission. 
Formed in the shock region, 
this emission may display velocity variations defined by the 
structure of the 
circumstellar gas, and thus one expects to observe phase-dependent 
variations in the line profile.

Thaller (1997) conducted a search for H$\alpha$ emission in
a large number of binaries with O-type components, 
including HD~165052, for which she found no sign of emission or visible 
distortion in the H$\alpha$ line profile.

We obtained five high-resolution spectra
centered at H$\alpha$ in different orbital phases.
We measured the equivalent widths of the H$\alpha$ line
profile finding no significant variations. 
Consequently, we are unable to confirm the existence 
of phase-related changes in this line.
To draw a conclusion, it will be necessary to expand the available data 
set with spectra corresponding to several distinct orbits, in order to
be able to appreciate possible orbit-to-orbit profile variations.
Also, as was pointed out by Thaller (1997), there is little expectation to
detect H$\alpha$ line-profile variations, as there seems to be a strong 
correlation between the emission and the evolutionary phase of the primary 
star, this phenomenon being very rare in systems with main-sequence 
components.  

From the minimum masses quoted in Table~\ref{orb_sol}, we can infer that
the orbital inclination is low, and thus the colliding wind effect 
may not be conspicuous.

\subsection{The X-ray lightcurve of HD 165052}

In a massive binary, X-ray emission is observed as a natural consequence of 
wind-wind interaction that produces a hot shocked gas region with 
temperatures in excess of a million degrees.
The density and relative velocity of the winds in the collision region 
and the wind absorption along the line of sight all vary as the stars revolve 
in their orbits. Depending directly on those parameters, 
one expects a phase-dependent change in the observed X-ray emission 
(Corcoran 1996 and references therein).

An X-ray light curve of HD165052 based on pointed observations with the
ROSAT Position Sensitive Proportional Counter (PSPC) was obtained 
by Corcoran (1996), 
and although this lightcurve was constructed using the ephemerides 
from Morrison \& Conti (1978), it shows a significant variability apparently 
tied to orbital phase. 

Using our new determination of the period, we phased the ROSAT PSPC 
observations in order to re-examine the X-ray light curve of HD~165052. 
\input{xobs.tex}

\begin{figure}
\includegraphics[width=70mm, bb= 130 240 470 540]{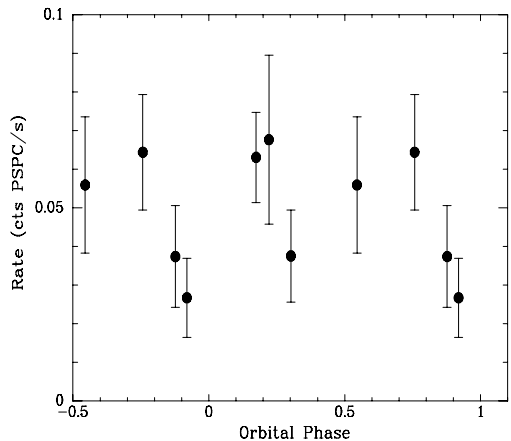}
\caption{X-ray ligth curve of HD~165052.}
\label{xcurv}
\end{figure}

As seen in Figure \ref{xcurv}, the light curve thus obtained shows 
clear phase-locked variations consisting of a rise of the observed X-ray flux 
near both quadratures, i.e., $\phi=0.22$ and 
$\phi=0.72$ approximately.

This twin-peaked structure of the X-ray emission is entirely consistent 
with the theoretical model developed by Pittard \& Stevens (1997) 
of colliding winds in an O+O binary system, in which the winds of each star
have similar momentum.
They explained the variation in the observed X-ray flux 
mainly as the result of changing absorption along the line of sight 
suffered by the intrinsic X-ray luminosity at different phases of the 
binary motion.
The occultation of the shocked region
by the stars has a minor contribution.
We therefore conclude that the effect of colliding winds is significant 
in this system, although attenuated by the unfavorable orbital inclination.

\subsection{The Struve-Sahade Effect}

The Struve-Sahade (S-S) effect is observed in many massive 
double-lined binaries
and consists in the apparent strengthening of the lines of the secondary
during orbital phases in which this component is approaching.
Reported for the first time by Bailey (1896) in the lines of $\mu_1$ Sco,
this phenomenon was noted and studied in a number of hot binaries by 
Sahade (1962), Bagnuolo et al. (1997), Howarth et al. (1997) among other
investigators.

\begin{figure}
\includegraphics[width=80mm, bb= 80 240 430 530]{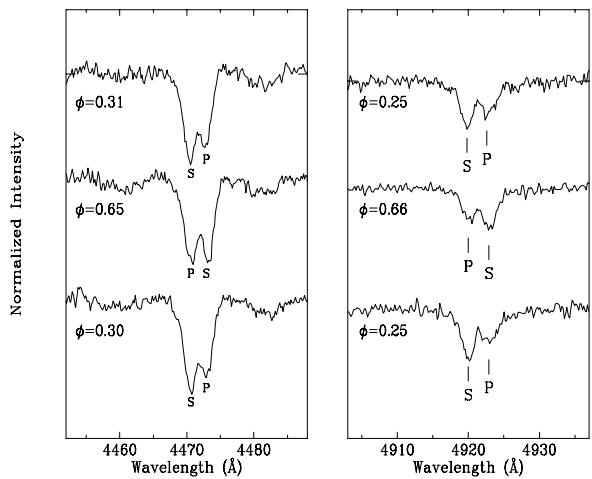}
\caption{Rectified spectrograms of HD~165052 at three orbital phases 
illustrating the Struve-Sahade effect observed in the He\,{\sc i} 4471\AA\ 
(left-hand panel) and 4921 \AA\ (right-hand panel) absorption lines.
``P'' and ``S'' tick marks show the primary and secondary components
respectively.}   
\label{efectoss}
\end{figure}

The origin of the Struve-Sahade effect still remains unclear, since 
there are a variety of mechanisms that can result in the variations of 
the line intensities of both binary components as function of the orbital 
phase. In O+O binary systems expected to contain colliding stellar winds, in 
accordance with the model proposed by Gies, Bagnuolo \& Penny (1997),
the S-S effect may be explained by a mechanism of 
localized photospheric heating induced by the wind-collision.
If the wind momentum of the primary dominates, the colliding-wind bow shock 
may reach close to the photosphere of the secondary 
and the X-ray flux arising from the shock region 
will preferentially heat one of the hemispheres of this star. 
Because of the Coriolis deflection due to orbital motion, the heated surface 
is best seen when the secondary is moving toward the observer. 
Thus, the contribution of the secondary to the binary spectrum is increased, 
yielding deeper secondary lines during the approaching phases.

As pointed out above, with a short period and very luminous components, 
HD~165052 appears as a potential candidate for colliding stellar winds, 
an idea strongly reinforced by the properties of the observed X-ray emission. 
Furthermore, from the ratio of the equivalent widths of well 
separated lines in the binary components, we estimated the visual 
luminosity  ratio of the system through Petrie's method (Petrie 1940).
Using for comparison the equivalent widths from Mathys (1988) 
corresponding to single stars of the same spectral types, we obtained
from the He\,{\sc ii}~4542\,\AA\ and He\,{\sc ii}~4686\,\AA\ lines,
$L($O6.5\,V$)/L($O7.5\,V$) \sim 0.79 \pm 0.05$, indicating  that the secondary 
has weaker wind than the primary.
In view of this, the possibility of detecting the S-S effect in the 
binary spectrum arises naturally (in spite of the relatively 
low orbital inclination).
We therefore examined our high-resolution 
spectra in detail, carefully comparing one quadrature with the other in order to detect 
systematic differences in the relative strength of the lines 
observed in the approaching and the receding phases.

We found variable line strength in several He\,{\sc i} absorption lines.
Figure~\ref{efectoss} displays normalized spectrograms around the He\,{\sc i}
4471\,\AA\ and 4922\,\AA\  absorption profiles at 
different orbital phases, where the effect can be appreciated. 
In fact, for example, in He\,{\sc i} 4471\,\AA, the secondary-to-primary line 
depth ratio, which is near unity for orbital phase $\phi\sim0.65$, appears 
to increase to a value significantly larger for phase $\phi\sim0.30$.
This absorption-line behavior provides an illustration of 
the S-S effect and consequently evidence supporting  
photospheric heating by the action
of colliding winds in HD~165052.

\section{Summary and Conclusions}

In this paper we have presented a detailed study of the O-type binary system 
HD~165052 based on recent high-resolution CCD spectroscopic observations.

We have re-determined the spectral types of both binary 
components, resulting in the spectral classification of
O6.5\,V + O7.5\,V. 
We have also detected the C\,{\sc iii} $\lambda5696$ emission in both spectral 
components, this line being stronger in the primary.

We have determined  an improved set of orbital elements 
using our high-resolution optical spectra. We thus found a 
slightly eccentric orbit 
($e=0.09$) with a period of 2.95510 d, which is consistent with the 
previous determination from archival {\em IUE} observations performed by  
Stickland et al. (1997). 
We obtained for the binary components velocity semi-amplitudes 
of 94.8 $\pm$ 0.5 km\,s$^{-1}$ and 104.7 $\pm$ 0.5 km\,s$^{-1}$, 
resulting in a mass-ratio $Q=0.9$.
Significant evidence for apsidal motion in HD~165052 was also presented. 

From the new orbital solution,  together with the projected rotational 
velocities of the components derived by Stickland et al (1997) and the
fact that both stars belong to luminosity class V, we showed that the
system has not reached synchronous rotation.

Assuming $M_V$ corresponding to the derived spectral types 
and luminosity class
for the binary components of HD~165052, one finds that
 the reddening law in the region should be normal, in contrast with
previous determinations. 

Finally, we investigated the interaction of winds in HD~165052 
through examination of a number of effects that provide evidence for 
colliding winds in close binary systems.
We found no emission or visible distortion of the H$\alpha$ profile,
but we certainly confirmed the presence of phase-locked variations of
the X-ray emission. The X-ray light curve obtained shows a twin-peaked 
structure 
with two maxima, one near each quadrature phase, in good agreement with 
current models for O+O systems with equal stellar winds 
(Pittard \& Stevens 1997).
We also detected the presence of the 
Struve-Sahade effect in the He\,{\sc i} absorption lines, reinforcing the idea 
that colliding winds could be in fact significant in HD~165052.

\section*{acknowledgements}

We gratefully thank Nolan Walborn for kindly providing the photographic
material analysed in this study, 
as well as for many useful suggestions and for
improving the English in an earlier version of this paper.

We acknowledge use at CASLEO of the CCD and data acquisition
system supported under US NSF grant AST-90-15827 to R.M. Rich, and want to 
thank the director and staff of CASLEO for the use of their facilities
and kind hospitality during the observing runs.

RHB acknowledges financial support from Fundaci\'on Antorchas
(Project No. 13783-5).

We want to thank our referee, Douglas Gies, for helpful suggestions.

\end{document}

%% file: runs.tex
\begin{table}
\centering
\caption{Instrumental configurations for different observing runs.}
\label{runs.tex}
\begin{tabular}[h]{@{}rccclr}
\hline
Id. & Date & Spectrograph  & Spectral    & S/N &  No. \\
    &      &               & Range [\AA] &     & Obs.\\
\hline
1 & 1994 June   & B\&C  & 3900-4700  &100& 17 \\   
2 & 1995 May    & REOSC-CD & 3750--7600 &140&5 \\ 
3 & 1995 Aug & REOSC-CD & 3750--7600 &140& 7 \\ 
4 & 1996 June   & REOSC-CD & 3750--7600 &140& 6 \\
5 & 1996 July   & REOSC-SD & 3950--7000 &320& 2 \\
6 & 1996 Aug & REOSC-CD & 3750--7600 &140& 4 \\
7 & 1996 Sep & REOSC-CD & 3750--7600 &140& 2 \\
8 & 1997 Aug & REOSC-CD & 3750--7600 &140& 2 \\
9 & 2000 June & REOSC-CD & 3750--7600 &140&12 \\
  &           & REOSC-SD & 3950--7000 &320& 4\\
10& 2001 June & REOSC-CD & 3300--6000 &140&12 \\
\hline		
\end{tabular}
\end{table}

%% file: radvel_all.tex
\begin{table*}
\centering
\hbox{
\rotcaption{High-resolution radial velocity measurements for HD~165052.}
\label{table2}
\begin{sideways}
\begin{tabular}{
r@{\hspace{3mm}}
r@{\hspace{3mm}}
r@{\hspace{3mm}}
r@{\hspace{3mm}}
r@{\hspace{3mm}}
r@{\hspace{3mm}}
r@{\hspace{3mm}}
r@{\hspace{3mm}}
r@{\hspace{3mm}}
r@{\hspace{3mm}}
r@{\hspace{3mm}}
r@{\hspace{3mm}}
r@{\hspace{3mm}}
r@{\hspace{3mm}}
r@{\hspace{3mm}}
r@{\hspace{3mm}}
r@{\hspace{3mm}}
r@{\hspace{3mm}}
r@{\hspace{3mm}}
r@{\hspace{3mm}}
}
\hline \\[-5pt]
HJD & He\,{\sc i} & He\,{\sc i} & Si\,{\sc iv} & He\,{\sc ii} & He\,{\sc i} & He\,{\sc i} & Mg\,{\sc ii} & He\,{\sc ii} & He\,{\sc ii} & He\,{\sc i} & He\,{\sc i} & He\,{\sc i} & He\,{\sc ii} & O\,{\sc iii} & C\,{\sc iv} &  C\,{\sc iv} & He\,{\sc i} & $V_R$ & $\sigma$ \\
$2\,400\,000+$ & 3819 & 4026 & 4088 & 4200 & 4387 & 4471 & 4481 & 4542 & 4686 & 4713 & 4921 & 5015 & 5411 & 5592 & 5801 &  5812 & 5875 & km\,s$^{-1}$ & km\,s$^{-1}$ \\
\hline\\
$\ast$$49852.677$&$	 -	$&$-	$&$-	$&$-	$&$-	$&$-21.7	$&$-	$&$-8.0	$&$-	 $&$-9.9	$&$-	$&$+3.1	 $&$+5.4	$&$-0.9	$&$-	$&$-	$&$-	$&$-5.3	$&$10.0	$\\[0.5ex]
$\ast$$49853.672$&$	 -	$&$-	$&$-	$&$-	$&$-	$&$+5.6	        $&$-	$&$-	$&$+10.4 $&$+7.2	$&$-	$&$+17.3 $&$+8.3	$&$-	$&$-	$&$-	$&$-	$&$+9.8	$&$4.5	$\\[0.5ex]
$49854.640$&$	 -	$&$+78.0$&$-	$&$-	$&$-	$&$+94.9	$&$-	$&$-	$&$-	$&$+90.6	$&$-	$&$-	$&$+89.5	$&$-	$&$-	$&$-	$&$-	$&$+88.2	$&$7.2	$\\
$         $&$	 -      $&$ -95.5 $&$-      $&$-      $&$-      $&$-99.6  $&$-      $&$ -     $&$-      $&$-98.4  $&$-      $&$-      $&$-100.7 $&$-      $&$-      $&$-      $&$-      $&$-98.5  $&$ 2.2	$\\[0.5ex]
$\ast$$49855.651$&$	 -	$&$-	$&$+11.9	$&$-	$&$-	$&$+5.5	$&$-	$&$+4.8	$&$-7.7	$&$-	$&$+13.9	$&$-	$&$- 	$&$-   	$&$-	$&$- 	$&$-	$&$+5.7	$&$8.5	$\\[0.5ex]
$49931.634$&$	 -	$&$+61.2	$&$+75.2	$&$-	$&$-	$&$+66.3	$&$-	$&$-	$&$-	$&$-	$&$-	$&$+71.0	$&$-	$&$+69.0	$&$+59.1	$&$+71.9	$&$+80.1	$&$+69.2	$&$6.9	$\\
$         $&$	 -      $&$ -87.0   $&$-90.6  $&$-      $&$-      $&$-89.4  $&$-      $&$-      $&$-      $&$-      $&$-      $&$-107.9 $&$-      $&$-82.8  $&$-96.5  $&$-99.0    $&$-78.2  $&$-91.4  $&$ 9.5 	$\\[0.5ex]
$49932.663$&$	 -80.7	$&$-76.3	$&$-73.4	$&$-	$&$-	$&$-75.1	$&$-	$&$-	$&$-	$&$-	$&$-	$&$-	$&$-	$&$-70.4	$&$-83.0	$&$-83.0	$&$-70.5	$&$-76.5	$&$5.2	$\\
$ 	 $&$	 +100.0	$&$+81.7	$&$+107.0	$&$-	$&$-	$&$+92.6	$&$-	$&$-	$&$-	$&$-	$&$-	$&$-	$&$-	$&$+95.3	$&$+87.1	$&$+96.7	$&$+103.1	$&$+95.4	$&$8.3	$\\[0.5ex]
$\ast$$49933.581$&$	 -2.3	$&$-10.9	$&$+5.6	$&$-	$&$-	$&$-0.1	$&$-	$&$-	$&$-	$&$-	$&$-	$&$-	$&$-	$&$-1.4	$&$-4.6	$&$0.0	$&$+2.8	$&$-1.4	$&$4.9	$\\[0.5ex]
$49934.577$&$	 -	$&$+66.2	$&$+81.6	$&$-	$&$-	$&$+78.0	$&$-	$&$-	$&$-	$&$-	$&$-	$&$-	$&$-	$&$+74.4	$&$+64.8	$&$+78.1	$&$+83.6	$&$+75.2	$&$7.3	$\\
$         $&$	 -      $&$ -84.3 $&$-76.9  $&$-      $&$-      $&$-80.3  $&$-      $&$-      $&$-      $&$-      $&$-      $&$-      $&$-      $&$-84.3  $&$-88.2  $&$-86.6  $&$-79.2  $&$-82.8  $&$ 4.1 	$\\[0.5ex]
$49935.605$&$	 -75.4	$&$-79.5	$&$-	$&$-	$&$-	$&$-73.3	$&$-	$&$-	$&$-76.3	$&$-	$&$-	$&$-	$&$-	$&$-80.5	$&$-80.9	$&$-80.8	$&$-74.5	$&$-77.6	$&$3.1	$\\
$	 $&$	 +86.4	$&$+101.5	$&$-	$&$-	$&$-	$&$+90.3	$&$-	$&$-	$&$+80.5	$&$-	$&$-	$&$-	$&$-	$&$+89.5	$&$+77.8	$&$+91.0	$&$+91.4	$&$+88.5	$&$7.3	$\\[0.5ex]
$\ast$$49936.657$&$	 +3.4	$&$-	$&$-2.3	$&$+6.7	$&$-	$&$-6.1	$&$-	$&$-	$&$-0.1	$&$-	$&$-	$&$-11.7	$&$+7.6	$&$-	$&$-4.9	$&$+3.3	$&$+4.6	$&$0.0	$&$6.2	$\\[0.5ex]
$49937.587$&$	 -	$&$-	$&$+76.0	$&$-	$&$-	$&$+69.4	$&$-	$&$-	$&$-	$&$-	$&$-	$&$+71.2	$&$-	$&$+76.4	$&$+64.7	$&$+67.6	$&$+78.4	$&$+71.9	$&$5.1	$\\
$         $&$	 -      $&$ -     $&$-76.5  $&$-      $&$-      $&$-72.0    $&$-      $&$-      $&$-      $&$-      $&$-      $&$-75.9  $&$-      $&$-74.8  $&$-72.8  $&$-70.1  $&$-70.6  $&$-73.2  $&$ 2.5 	$\\[0.5ex]
$50239.906$&$	 -	$&$-	$&$-	$&$-	$&$-	$&$-74.9	$&$-	$&$-	$&$-	$&$-	$&$-	$&$-	$&$-	$&$-	$&$-72.1	$&$-77.1	$&$-72.1	$&$-74.0	$&$2.4	$\\
$	 $&$	 -	$&$-	$&$-	$&$-	$&$-	$&$+72.7	$&$-	$&$-	$&$-	$&$-	$&$-	$&$-	$&$-	$&$-	$&$+84.1	$&$+75.7	$&$+74.9	$&$+76.8	$&$5.0	$\\[0.5ex]
$\ast$$50240.883$&$	 -	$&$-0.5	$&$+12.6	$&$-	$&$-	$&$+2.5	$&$-	$&$-	$&$-	$&$-	$&$-	$&$+3.1	$&$+3.0	$&$-0.4	$&$-3.7	$&$+2.5	$&$+1.0	$&$+2.2	$&$4.5	$\\[0.5ex]
$50241.823$&$	 -	$&$-	$&$-	$&$-	$&$-	$&$+88.0	$&$-	$&$-	$&$-	$&$-	$&$-	$&$+76.3	$&$-	$&$+81.8	$&$+75.5	$&$+86.8	$&$+86.9	$&$+82.5	$&$5.6	$\\
$         $&$	 -      $&$ -     $&$-      $&$-      $&$-      $&$-79.7  $&$-      $&$-      $&$-      $&$-      $&$-      $&$-81.5  $&$-      $&$-98.3  $&$-94.5  $&$-101.0   $&$-90.7  $&$-90.9  $&$ 8.7	$\\[0.5ex]
$\ast$$50243.894$&$	 -0.5	$&$+3.1	$&$+8.0	$&$-	$&$-	$&$+0.9	$&$-	$&$-	$&$+2.8	$&$-	$&$-	$&$+1.6	$&$-	$&$-4.0	$&$-4.1	$&$-5.7	$&$+3.6	$&$+0.5	$&$4.2	$\\[0.5ex]
$50244.902$&$	 -	$&$-	$&$-	$&$-	$&$-	$&$+74.2	$&$-	$&$-	$&$-	$&$+83.1	$&$-	$&$+75.9	$&$-	$&$+79.1	$&$+65.6	$&$+58.3	$&$+72.5	$&$+72.6   $&$ 8.6	$\\
$         $&$	 -      $&$ -     $&$-      $&$-      $&$-      $&$-69.9  $&$-      $&$-      $&$-      $&$-79.1  $&$-      $&$-79.5  $&$-      $&$-75.8  $&$-72.9  $&$-80.5  $&$-78.0    $&$-76.5  $&$3.9 	$\\[0.5ex]
$50245.857$&$	 -87.9	$&$-59.5	$&$-	$&$-	$&$-	$&$-74.3	$&$-	$&$-	$&$-	$&$-	$&$-	$&$-75.7	$&$-	$&$-76.2	$&$-69.7	$&$-	$&$-80.0	$&$-74.7	$&$8.8	$\\
$	 $&$	 +78.6	$&$+93.9	$&$-	$&$-	$&$-	$&$+75.2	$&$-	$&$-	$&$-	$&$-	$&$-	$&$+79.7	$&$-	$&$+84.6	$&$+80.7	$&$-	$&$+84.8	$&$+82.5	$&$6.0	$\\[0.5ex]
$50293.763$&$	 -86.2	$&$-	$&$-72.6	$&$-	$&$-	$&$-69.5	$&$-	$&$-	$&$-	$&$-64.7	$&$-	$&$-72.1	$&$-	$&$-78.0	$&$-76.9	$&$-81.4	$&$-69.0	$&$-74.5	$&$6.7	$\\
$	 $&$	 +70.8	$&$-	$&$+89.5	$&$-	$&$-	$&$+87.4	$&$-	$&$-	$&$-	$&$+92.5	$&$-	$&$+86.3	$&$-	$&$+80.7	$&$+72.5	$&$+73.4	$&$+89.8	$&$+82.5	$&$8.4	$\\[0.5ex]
$\ast$$50295.770 $&$	 +6.4	$&$+12.5	$&$-	$&$-	$&$-	$&$+4.1	$&$-	$&$-	$&$-	$&$-	$&$-	$&$+8.4	$&$-	$&$-	$&$-	$&$-	$&$+4.2	$&$+7.1	$&$3.5	$\\[0.5ex]
$50296.787$&$	 -	$&$-	$&$-	$&$-	$&$-	$&$-	$&$-	$&$-	$&$-	$&$-	$&$-	$&$-	$&$-	$&$-66.4	$&$-66.2	$&$-66.1	$&$-54.3	$&$-63.2	$&$5.9	$\\
$	 $&$	 -	$&$-	$&$-	$&$-	$&$-	$&$-	$&$-	$&$-	$&$-	$&$-	$&$-	$&$-	$&$-	$&$+62.3	$&$+63.5	$&$+66.5	$&$+65.1	$&$+64.3	$&$1.8	$\\[0.5ex]
$\ast$$50298.730 $&$	 -	$&$-	$&$-	$&$-	$&$-	$&$+0.4	$&$-	$&$-	$&$-	$&$+16.7	$&$-	$&$+7.9	$&$-	$&$-	$&$-	$&$-	$&$+18.9	$&$+11.0	$&$8.5	$\\[0.5ex]
$\ast$$50348.614$&$	 -8.4	$&$-3.3	$&$-0.9	$&$-	$&$-5.9	$&$+0.4	$&$-	$&$-	$&$-	$&$-	$&$-12.2	$&$-1.6	$&$-	$&$-6.9	$&$-0.5	$&$-14.0	$&$-2.5	$&$-5.1	$&$4.8	$\\[0.5ex]
$50671.471$&$	 -73.1	$&$-78.9	$&$-70.9	$&$-	$&$-	$&$-75.0	$&$-	$&$-	$&$-75.4	$&$-	$&$-	$&$-81.0	$&$-	$&$-	$&$-83.1	$&$-87.6	$&$-	$&$-78.1	$&$5.6	$\\
$	 $&$	 +92.5	$&$+82.6	$&$+94.0	$&$-	$&$-	$&$+102.6	$&$-	$&$-	$&$+96.0	$&$-	$&$-	$&$+98.2	$&$-	$&$-	$&$+87.4	$&$+99.1	$&$-	$&$+94.0	$&$6.5	$\\[0.5ex]
$\ast$$50672.494$&$	 -	$&$-	$&$-	$&$-	$&$-	$&$-	$&$-	$&$-	$&$-	$&$-	$&$-	$&$-14.5	$&$-0.5	$&$-2.1	$&$+0.2	$&$-12.1	$&$-0.5	$&$-4.9	$&$6.6	$\\[0.5ex]
$\ast$$51713.844$&$	 -1.0	$&$-23.4	$&$-5.0	$&$-	$&$-	$&$-3.3	$&$-	$&$-	$&$-	$&$-	$&$-	$&$-12.8	$&$-	$&$-1.9	$&$+4.4	$&$-5.4	$&$+0.7	$&$-5.3	$&$8.3	$\\[0.5ex]
$51714.835$&$	 -	$&$-86.6	$&$-86.9	$&$-	$&$-	$&$-83.0	$&$-	$&$-	$&$-	$&$-	$&$-	$&$-82.7	$&$-	$&$-	$&$-91.2	$&$-92.6	$&$-87.5	$&$-87.2	$&$3.7	$\\
$	 $&$	 -	$&$+110.1	$&$+102.3	$&$-	$&$-	$&$+104.8	$&$-	$&$-	$&$-	$&$-	$&$-	$&$+105.0	$&$-	$&$-	$&$+94.9	$&$+99.9	$&$+103.4	$&$+102.9	$&$4.7	$\\[0.5ex]
\hline\\
\end{tabular}
\end{sideways}
}
\end{table*}

\setcounter{table}{2}

\begin{table*}
\centering
\hbox{
\begin{sideways}
\begin{tabular}{
r@{\hspace{2.5mm}}
r@{\hspace{2.5mm}}
r@{\hspace{2.5mm}}
r@{\hspace{2.5mm}}
r@{\hspace{2.5mm}}
r@{\hspace{2.5mm}}
r@{\hspace{2.5mm}}
r@{\hspace{2.5mm}}
r@{\hspace{2.5mm}}
r@{\hspace{2.5mm}}
r@{\hspace{2.5mm}}
r@{\hspace{2.5mm}}
r@{\hspace{2.5mm}}
r@{\hspace{2.5mm}}
r@{\hspace{2.5mm}}
r@{\hspace{2.5mm}}
r@{\hspace{2.5mm}}
r@{\hspace{2.5mm}}
r@{\hspace{2.5mm}}
r@{\hspace{2.5mm}}
}
\multicolumn{20}{l}{Table 2 (continued)}\\[1ex] 
\hline\\[-5pt]
HJD & He\,{\sc i} & He\,{\sc i} & Si\,{\sc iv} & He\,{\sc ii} & He\,{\sc i} & He\,{\sc i} & Mg\,{\sc ii} & He\,{\sc ii} & He\,{\sc ii} & He\,{\sc i} & He\,{\sc i} & He\,{\sc i} & He\,{\sc ii} & O\,{\sc iii} & C\,{\sc iv} &  C\,{\sc iv} & He\,{\sc i} & $V_R$ & $\sigma$ \\
$2\,400\,000+$ & 3819 & 4026 & 4088 & 4200 & 4387 & 4471 & 4481 & 4542 & 4686 & 4713 & 4921 & 5015 & 5411 & 5592 & 5801 & 5812 & 5875 & km\,s$^{-1}$ & km\,s$^{-1}$ \\
\hline\\
$51714.862$&$	 -	$&$-	$&$-	$&$-	$&$-	$&$-	$&$-	$&$-	$&$-	$&$-	$&$-	$&$-	$&$-91.5	$&$-106.0	$&$-88.7	$&$-95.2	$&$-86.0	$&$-93.5	$&$7.8	$\\
$	 $&$	 -	$&$-	$&$-	$&$-	$&$-	$&$-	$&$-	$&$-	$&$-	$&$-	$&$-	$&$-	$&$+80.3	$&$+110.0	$&$+98.5	$&$+104.0	$&$+103.4	$&$+99.2	$&$11.3	$\\[0.5ex]
$\ast$$51715.548$&$	 -1.2	$&$-4.4	$&$+6.4	$&$-	$&$-	$&$+3.8	$&$-	$&$-	$&$-	$&$-	$&$+6.5	$&$+5.4	$&$+0.8	$&$-0.1	$&$-4.2	$&$-4.7	$&$3.5	$&$+1.1	$&$4.3	$\\[0.5ex]
$\ast$$51715.737$&$	 -	$&$-	$&$-	$&$-	$&$-	$&$-	$&$-	$&$-	$&$-	$&$-	$&$-	$&$-	$&$+1.0	$&$+0.8	$&$-1.6	$&$+6.7	$&$+1.0	$&$+1.6	$&$3.1	$\\[0.5ex]
$51715.796$&$	 -	$&$-	$&$-	$&$-	$&$-	$&$-	$&$-	$&$-	$&$-	$&$-	$&$-	$&$+64.6	$&$-	$&$+66.9	$&$+54.2	$&$+55.1	$&$+65.2	$&$+61.2	$&$6.0	$\\
$         $&$	 -      $&$ -     $&$-      $&$-      $&$-      $&$ -     $&$-      $&$-      $&$-      $&$-      $&$-      $&$-58.0    $&$-      $&$-64.5  $&$-62.7  $&$-67.2  $&$-61.9  $&$-62.9  $&$ 3.4 	$\\[0.5ex]
$51716.570 $&$	 +72.0	$&$-	$&$+74.8	$&$-	$&$-	$&$+79.0	$&$-	$&$-	$&$-	$&$-	$&$+75.4	$&$+73.0	$&$-	$&$+69.6	$&$+65.6	$&$+66.0	$&$+69.4	$&$+71.6	$&$4.4	$\\
$         $&$	 -75.6  $&$ -     $&$-67.4  $&$-      $&$-      $&$-70.0    $&$-      $&$-      $&$-      $&$-      $&$-61.7  $&$-71.7  $&$-      $&$-69.3  $&$-66.7  $&$-68.1  $&$-66.4  $&$-68.5  $&$ 3.8 	$\\[0.5ex]
$51716.656$&$	 -	$&$-	$&$-	$&$-	$&$-	$&$-	$&$-	$&$-	$&$-	$&$-	$&$-	$&$-	$&$-	$&$+38.8	$&$+39.5	$&$-	$&$+23.7	$&$+34.0	$&$8.9	$\\
$         $&$	 -      $&$ -     $&$-      $&$-      $&$-      $&$-      $&$-      $&$-      $&$-      $&$-      $&$-      $&$-      $&$-      $&$-35.1  $&$-40.5  $&$-      $&$-50.2  $&$-41.9  $&$ 7.6	$\\[0.5ex]
$\ast$$51716.778$&$	 -18.1	$&$-15.2	$&$-4.2	$&$-	$&$-	$&$-10.1	$&$-	$&$-	$&$-	$&$-13.1	$&$-	$&$-12.4	$&$-	$&$-	$&$-14.0	$&$-1.6	$&$-11.8	$&$-11.2	$&$5.2	$\\[0.5ex]
$51717.609$&$	 -68.4	$&$-83.0	$&$-79.0	$&$-	$&$-	$&$-77.2	$&$-	$&$-	$&$-85.5	$&$-	$&$-	$&$-85.4	$&$-	$&$-83.2	$&$-	$&$-	$&$-82.7	$&$-80.5	$&$5.7	$\\
$	 $&$	 +94.2	$&$+101.7	$&$+103.6	$&$-	$&$-	$&$+95.8	$&$-	$&$-	$&$+93.7	$&$-	$&$-	$&$+92.2	$&$-	$&$+91.6	$&$+91.4	$&$-	$&$+96.4	$&$+95.6	$&$4.4	$\\[0.5ex]
$51717.727$&$	 -	$&$-	$&$-	$&$-	$&$-	$&$-	$&$-	$&$-	$&$-	$&$-	$&$-	$&$-	$&$-85.4	$&$-95.9	$&$-91.0	$&$-	$&$-89.5	$&$-90.5	$&$4.3	$\\
$	 $&$	 -	$&$-	$&$-	$&$-	$&$-	$&$-	$&$-	$&$-	$&$-	$&$-	$&$-	$&$-	$&$+83.0	$&$+97.2	$&$+100.9	$&$-	$&$+100.1	$&$+95.3	$&$8.3	$\\[0.5ex]
$51717.829$&$	 -99.6	$&$-99.3	$&$-	$&$-	$&$-	$&$-85.8	$&$-	$&$-	$&$-	$&$-97.7	$&$-	$&$-84.8	$&$-	$&$-89.2	$&$-90.6	$&$-96.7	$&$-93.6	$&$-93.0	$&$5.7	$\\
$	 $&$	 +90.4	$&$+90.0	$&$-	$&$-	$&$-	$&$+102.0	$&$-	$&$-	$&$-	$&$+100.8	$&$-	$&$+98.8	$&$-	$&$+98.0	$&$+97.1	$&$+101.0	$&$+101.4	$&$+97.7	$&$4.6	$\\[0.5ex]
$52066.791$&$	 -	$&$-71.1	$&$-64.2	$&$-67.8	$&$-	$&$-59.0	$&$-	$&$-	$&$-	$&$-	$&$-75.5	$&$-87.5	$&$-64.9	$&$-65.2	$&$-	$&$-	$&$-75.4	$&$-70.1	$&$8.5	$\\
$	 $&$	 -	$&$+83.9	$&$+106.0	$&$-	$&$-	$&$+94.0	$&$-	$&$-	$&$-	$&$-	$&$+83.5	$&$+107.5	$&$+87.0	$&$+83.0	$&$-	$&$-	$&$+79.9	$&$+90.6	$&$10.8	$\\[0.5ex]
$52067.768$&$	 -	$&$+81.3	$&$-	$&$-	$&$-	$&$+103.0	$&$-	$&$-	$&$-	$&$-	$&$-	$&$-	$&$-	$&$-	$&$-	$&$-	$&$+102.5	$&$+95.6	$&$12.4	$\\
$         $&$	 -      $&$ -109.5$&$  -    $&$-      $&$-      $&$-109.0   $&$-      $&$-      $&$-      $&$-      $&$-      $&$-      $&$-      $&$-      $&$-      $&$-      $&$-113.0   $&$-110.5 $&$ 2.2	$\\[0.5ex]
$52069.730 $&$	 -69.0	$&$-73.5	$&$-78.4	$&$-70.6	$&$-	$&$-	$&$-75.2	$&$-	$&$-	$&$-	$&$-69.4	$&$-81.1	$&$-	$&$-85.1	$&$-	$&$-	$&$-76.0	$&$-75.4	$&$5.8	$\\
$	 $&$	 +76.1 	$&$+65.6 	$&$+96.7 	$&$+114.3 	$&$-	$&$+87.4 	$&$+87.8 	$&$-	$&$-	$&$-	$&$+76.2 	$&$+88.3 	$&$-	$&$-	$&$-	$&$-	$&$+86.9 	$&$+86.6	$&$13.8	$\\[0.5ex]
$52069.794$&$	 -67.0   	$&$-79.2 	$&$-70.0   	$&$-	$&$-62.4 	$&$-61.7 	$&$-	$&$-	$&$-	$&$-	$&$-	$&$-	$&$-56.4 	$&$-69.1 	$&$-	$&$-	$&$-72.6 	$&$-67.3	$&$7.1	$\\
$	 $&$	 +74.4	$&$+58.3	$&$+89.4	$&$-	$&$+107.1	$&$+68.6	$&$-	$&$-	$&$-	$&$-	$&$-	$&$-	$&$+72.4	$&$+67.0	$&$-	$&$-	$&$+72.1	$&$+76.1	$&$15.2	$\\[0.5ex]
$52070.621$&$	 -	$&$+101.1 	$&$+90.7 	$&$-	$&$+102.4 	$&$+96.6 	$&$+78.9 	$&$+91.7 	$&$+87.0 	$&$-	$&$-	$&$+92.9 	$&$+98.9 	$&$+93.9 	$&$+92.9 	$&$+92.7 	$&$+93.8 	$&$+93.3	$&$6.1	$\\
$         $&$	 -      $&$ -90.4 $&$-98.6  $&$-      $&$-106.2 $&$-95.3  $&$-99.8  $&$-81.4  $&$-92.9  $&$-      $&$-      $&$-110.4 $&$-93.5  $&$-99.7  $&$-97.4  $&$-      $&$-98.7  $&$-97.1  $&$ 7.4	$\\[0.5ex]
$52070.731$&$	 +101.6 	$&$+100.8 	$&$+110.2 	$&$-	$&$-	$&$+111.1 	$&$+88.7 	$&$-	$&$+95.7 	$&$+107.2 	$&$-	$&$-	$&$+106.3 	$&$+104.4 	$&$+109.1 	$&$-	$&$+103.7 	$&$+103.5	$&$6.7	$\\
$         $&$	 -      $&$-118.2 $&$-107.7	$&$-      $&$-      $&$-104.3 $&$-106.1 $&$-90.9  $&$-103.4 $&$-94.6  $&$-      $&$-      $&$-91.0  $&$-111.8 $&$-114.3 $&$-      $&$-111.7 $&$-104.9 $&$ 9.3	$\\[0.5ex]
$52070.792$&$	 -	$&$+92.4 	$&$+107.0 	$&$-	$&$-	$&$+110.2 	$&$+100.5 	$&$+95.2 	$&$+97.4 	$&$+117.8 	$&$+92.6 	$&$+105.0 	$&$+103.3 	$&$+104.7 	$&$-	$&$-	$&$+107.6 	$&$+102.8	$&$7.6	$\\
$	 $&$	 -	$&$-111.0 $&$-112.3 $&$-	$&$-	$&$-104.5 $&$-115.0 $&$-99.0 	$&$-106.8 $&$-96.0 	$&$-110.4 $&$-113.9 $&$-92.3 	$&$-115.8 $&$-	$&$-	$&$-111.4 $&$-107.3	$&$7.8	$\\[0.5ex]
$\ast$$52071.663$&$	 +2.3 	$&$-2.3 	$&$+2.7 	$&$+4.5 	$&$+6.0 	$&$+6.8 	$&$-1.3 	$&$+6.6 	$&$+4.4 	$&$+14.0 	$&$+5.4 	$&$+7.4 	$&$+2.8 	$&$-3.3 	$&$+0.2 	$&$-0.3 	$&$+4.9 	$&$+3.5	$&$4.2	$\\[0.5ex]
$\ast$$52071.746$&$	 +3.9	$&$+0.7	$&$+15.0	$&$+13.6	$&$+13.4	$&$+7.9	$&$+2.1	$&$+5.0	$&$+4.7	$&$+12.5	$&$+10.6	$&$+1.0	$&$+2.0	$&$+0.3	$&$+2.6	$&$+12.4	$&$+3.7	$&$+6.5	$&$5.2	$\\[0.5ex]
$\ast$$52071.810 $&$	 +7.4 	$&$-2.8 	$&$+16.0 	$&$+5.5 	$&$-1.2 	$&$+2.7 	$&$+5.5 	$&$+0.9	$&$+6.7 	$&$+5.5 	$&$+22.2 	$&$+19.2 	$&$+2.6 	$&$-9.0 	$&$-0.2 	$&$+7.6 	$&$+5.5 	$&$+5.5	$&$7.8	$\\[0.5ex]
$52072.663$&$	 -	$&$-71.3 	$&$-87.3 	$&$-	$&$-95.0 	$&$-67.5 	$&$-	$&$-	$&$-80.5 	$&$-63.4 	$&$-	$&$-79.2 	$&$-76.9 	$&$-62.7 	$&$-75.3 	$&$-70.0	$&$-74.8	$&$-75.3	$&$9.4	$\\
$	 $&$	 -	$&$+88.9 	$&$+98.4 	$&$-	$&$+93.0 	$&$+88.8 	$&$-	$&$-	$&$+77.6 	$&$-	$&$+101.2 	$&$+99.6 	$&$+90.6 	$&$+80.0 	$&$+90.2 	$&$+87.0 	$&$+86.1 	$&$+90.1	$&$7.2	$\\[0.5ex]
$52072.717$&$	 -67.9 	$&$-71.1 	$&$-	$&$-	$&$-	$&$-75.1 	$&$-61.2 	$&$-	$&$-	$&$-	$&$-	$&$-	$&$-65.3 	$&$-68.7 	$&$-71.2 	$&$-	$&$-75.0 	$&$-69.4	$&$4.7	$\\
$	 $&$	 -	$&$+80.2 	$&$-	$&$-	$&$-	$&$+73.0 	$&$-	$&$-	$&$-	$&$-	$&$-	$&$-	$&$+79.4 	$&$+77.7 	$&$+79.6 	$&$+86.2 	$&$+76.1 	$&$+78.9	$&$4.1	$\\[0.5ex]
\hline
\multicolumn{10}{l}{\footnotesize $\ast$ Value omitted from orbital solution}\\
\end{tabular}
\end{sideways}
}
\end{table*}

%% file: vr3.tex
\begin{table*}
\caption{Radial velocity measurements considered in the orbital solution
of HD~165052.}
\label{vr}
\begin{tabular}{@{}rrrrrrrr}
\hline\\[-5pt]
HJD           & Phase & V$_1$&  O-C &n$_1$& V$_2$& O-C&n$_2$ \\ 
$2\,400\,000+$& $\phi$& km\,s$^{-1}$ & km\,s$^{-1}$ & & 
km\,s$^{-1}$ & km\,s$^{-1}$ & \\ \hline \\
$49854.640$& $0.210$ & $+88.2$ & $-4.1$ & $4$ & $-98.5$ & $+1.3$ & $4$\\  
$49931.634$ & $0.265$ & $+69.2$ & $-5.9$ & $8$ & $-91.4$ & $-10.6$ & $8$\\ 
$49932.663  $&$ 0.613 $ & $-76.5  $&$ +1.7 $&$ 8 $&$ +95.4 $&$ +6.7  $& $8$\\ 
$49934.577  $&$ 0.261 $ & $ +75.2 $&$ -1.5 $&$ 7 $&$ -82.8 $&$ -0.3  $& $7$\\ 
$49935.605  $&$ 0.609 $ & $ -77.6 $&$ -0.4 $&$ 8 $&$  +88.5 $&$ +1.0 $&$8$\\ 
\\
$49937.587  $&$ 0.280 $ & $ +71.9 $&$ +2.6 $&$ 7 $&$  -73.2 $&$ +1.2 $&$ 7$\\ 
$50239.906  $&$ 0.583 $ & $ -74.0 $&$ -4.1 $&$ 4 $&$  +76.8 $&$ -2.7 $&$ 4$\\ 
$50241.823  $&$ 0.232 $ & $ +82.5 $&$ -3.9 $&$ 6 $&$  -90.9 $&$ +2.4 $&$ 6$\\ 
$50244.902  $&$ 0.274 $ & $ +72.6 $&$ +1.5 $&$ 7 $&$  -76.5 $&$ +0.2 $&$ 7$\\ 
$50245.857  $&$ 0.597 $ & $-74.7 $&$ -0.6  $&$ 7 $&$ +82.5  $&$ -1.6 $&$ 7$\\ 
\\
$50293.763  $&$ 0.809 $ & $-74.5 $&$ -4.6 $&$ 9 $&$ +82.6 $ &$ +3.1 $&$ 9$\\ 
$50296.787  $&$ 0.832 $ & $-63.2 $&$ -3.2 $&$ 4 $&$ +64.3  $&$ -4.2 $&$ 4$\\ 
$50671.471  $&$ 0.624 $ & $-78.1 $&$ +2.8 $&$ 8 $&$ +94.0  $&$ +2.4 $&$ 8$\\ 
$51714.835  $&$ 0.697 $ & $-87.2 $&$ +2.7 $&$ 7 $&$ +102.9 $&$ +1.3 $&$ 7$\\ 
$51714.862  $&$ 0.706 $ & $-93.5 $&$ -3.6 $&$ 5 $&$ +99.2  $&$ -2.4 $&$ 5$\\ 
\\
$51715.796  $&$ 0.022 $ & $+61.2 $&$ +0.4 $&$ 5 $&$ -62.9  $&$ +2.1 $&$ 5$\\ 
$51716.570  $&$ 0.284 $ & $+71.6 $&$ +4.1 $&$ 9 $&$ -68.5  $&$ +3.9 $&$ 9$\\ 
$\dagger51716.656$ & $0.313$ & $+34.0$ & $-$ & $3$ & $-41.9$ & $ - $&$ 3$\\ 
$51717.609  $&$ 0.635$ & $-80.7 $&$ +2.5 $&$ 9 $&$ +95.6 $ &$ +1.4 $&$ 9$\\ 
$51717.727  $&$ 0.675$ & $-90.5 $&$ -1.7 $&$ 4 $&$ +95.3  $&$ -5.0 $&$ 4$\\ 
\\
$51717.829  $&$ 0.710$ & $-93.0 $&$ -3.2 $&$ 9 $&$ +97.7  $&$ -3.8 $&$ 9$\\ 
$52066.791  $&$ 0.798$ & $-70.1 $&$ +3.7 $&$ 9 $&$ +90.6  $&$ +6.7 $&$ 8$\\
$52067.768  $&$ 0.129$ & $+95.6 $&$ -2.9 $&$ 3 $&$ -110.5 $&$ -3.8 $&$ 3$\\
$52069.730  $&$ 0.792$ & $-75.4 $&$ +0.3 $&$ 9 $&$ +86.6 $ &$ +0.7 $&$ 9$\\
$52069.794  $&$ 0.814$ & $-67.3 $&$ +0.5 $&$ 7 $&$ +76.2  $&$ -1.0 $&$ 7$\\
\\
$52070.621 $&$ 0.094$ & $+93.3 $ &$ +1.3 $&$ 12 $&$ -97.1  $&$ +2.4 $&$ 13 $\\
$52070.731 $&$ 0.131$ & $+103.5 $&$ +4.7 $&$ 11 $&$ -104.9 $&$ +2.0 $&$ 11 $\\
$52070.792 $&$ 0.152$ & $+102.8 $&$ +3.1 $&$ 12 $&$ -107.4 $&$ +0.5 $&$ 12 $\\
$52072.663 $&$ 0.785$ & $-75.3 $ &$ +2.7 $&$ 12 $&$ +90.1 $ &$ +1.7 $&$ 12 $\\
$52072.717 $&$ 0.803$ & $-69.4 $ &$ +2.6 $&$ 7  $&$ +78.9  $&$ -2.9 $&$ 7 $\\
\\
\hline
\multicolumn{8}{l}{\footnotesize $\dagger$ Value discarded from the final solution due to pair blending 
}\\
\end{tabular}
\end{table*}

%% file: orb_sol.tex
\begin{table}
\caption{Orbital elements of HD~165052 binary components derived from
radial velocities measured in high-resolution spectrograms}
\label{orb_sol}
\begin{center}
\begin{tabular}{@{}lr}
\hline
 &  \\
$P$~(days)                      & $2.95510 \pm 0.00001$    \\
\hline                             
 & \\
$e$                                 & $ 0.09 \pm 0.004$          \\
$\omega\,(^\circ)$               & $ 296.7 \pm 3.5  $        \\           
$T_o$~(HJD)                        &$ 2449871.75 \pm 0.03$       \\
$T\,V_{max}$~(HJD)                &$ 2449872.19\pm 0.03$       \\
$V_o$~(km\,s$^{-1}$)              & $ 1.05 \pm 0.31$          \\
                                  &                             \\
$K_1$~(km\,s$^{-1}$)                 & $ 94.8 \pm 0.5$      \\
$K_2$~(km\,s$^{-1}$)                 & $ 104.7 \pm 0.5$       \\
                                       &                         \\
$a_1 \sin i$~(R$_{\odot}$)        &$ 5.51 \pm 0.03 $        \\
$a_2 \sin i$~(R$_{\odot}$)        &$ 6.09 \pm 0.03 $        \\
                                       &                         \\
$M_1 \sin^3 i$~(M$_{\odot}$)   &$1.26 \pm 0.03$             \\
$M_2 \sin^3 i$~(M$_{\odot}$)  &$1.14 \pm 0.03$               \\
$Q~(M_2/M_1)$                    &$0.90 \pm 0.01$              \\
                                   &                              \\
Prob. error~(km\,s$^{-1}$)            & $2.21$                       \\  
                                   &                              \\
\hline
\end{tabular}
\end{center}
\end{table}

%% file: xobs.tex
\begin{table}
\caption{ROSAT PSPC observations of HD~165052. The values listed in the third 
column represent a simple average of all the observations obtained
during the same Julian day. The fourth column shows the corresponding 
errors. The orbital phases were computed with the ephemeris presented 
in Table~\ref{orb_sol}.} 
\label{xobs}
\begin{center}
\begin{tabular}{l c c r}
\hline
HJD      & Phase   &  Rate                 & $\sigma$   \\
2440000+ & $\phi$ & (PSPC counts s$^{-1}$) &            \\ 
\hline
          &         &           &            \\
8332.6553 &  0.174  & 0.063 &   0.011 \\
8900.1735 &  0.221  & 0.067 &   0.021 \\
8901.7591 &  0.757  & 0.064 &   0.014 \\
8904.0855 &  0.545  & 0.055 &   0.017 \\
9878.5524 &  0.302  & 0.037 &   0.011 \\
9079.4175 &  0.877  & 0.037 &   0.013 \\
9079.5420 &  0.919  & 0.026 &   0.010 \\
          &         &           &             \\
\hline
\end{tabular}
\end{center}
\end{table}